\documentclass{optica-article}

\journal{opticajournal} 

\articletype{Research Article}

\usepackage{graphicx} 
\usepackage{float} 
\usepackage{wrapfig} 
\usepackage[labelfont=bf]{caption} 
\usepackage{caption, subcaption} 

\usepackage{tikz,tikz-3dplot}
\usetikzlibrary{arrows.meta,bending,matrix,positioning,patterns,fit,angles,spy}

\usepackage{bm} 
\usepackage{hyperref} 
\usepackage{lineno} 

\begin{document}

\pagestyle{plain}  
\pagenumbering{arabic}  

\title{Inverse freeform design of a parallel-to-two-target reflector system}
\author{P. A. Braam,\authormark{1,*} J. H. M. ten Thije Boonkkamp,\authormark{1} M. J. H. Anthonissen,\authormark{1} K. Mitra,\authormark{1} R. Beltman,\authormark{2} and W. L. IJzerman\authormark{1,2}}
\address{\authormark{1}CASA, Department of Mathematics and Computer Science, Eindhoven University of Technology,\\ P.O. Box 513, 5600 MB Eindhoven, The Netherlands\\
\authormark{2}Signify Research, High Tech Campus 7, 5656 AE Eindhoven, The Netherlands}
\email{\authormark{*}p.a.braam@tue.nl}

\begin{abstract*} 
We present an inverse method for transforming a given parallel light emittance to two light distributions at different parallel target planes using two freeform reflectors. The reflectors control both the spatial and directional target coordinates of light rays. To determine the shape and position of the reflectors, we derive generating functions and use Jacobian equations to find the optical mappings to the two targets. The model is solved numerically by a three-stage least-squares algorithm. A feasibility condition is derived, which ensures that the reflectors are not self-intersecting. Several examples validate this condition and demonstrate the algorithm's capability of realizing complex distributions.
\end{abstract*}

\section{Introduction}
The goal in non-imaging optics is to design optical surfaces that convert a source light distribution to a target distribution. The two most commonly used methods are ray-trace methods and inverse methods. Ray-trace methods entail an iterative procedure where light rays propagate through the optical system and based on the deviation from the desired target distribution, the shape(s) of the optical surface(s) are altered \cite{RayT,Wang2,Tang}. Although these methods can be computationally expensive and often involve adjusting system parameters to gradually improve performance, recently developed methods have shown ways to improve the accuracy and reduction of computation time \cite{Wang1,Willem2024}. On the other hand, inverse methods provide a promising alternative approach by using the principles of geometrical optics and conservation of energy to compute the shapes of optical surfaces directly \cite{Martijn1}.

Inverse methods can be classified into two categories: methods that use optimization strategies and methods that directly solve a (generalized) Monge-Ampère equation or generated Jacobian equation; for a comprehensive overview, see \cite[p. 133]{Lotte}. Doskolovich \textit{et al.} used methods from the first category and solved an optimal transport problem by reducing it to a linear assignment problem \cite{Doskolovich}. Methods from the second category were used by Brix \textit{et al.} who solved the Monge-Ampère equation using a collocation method \cite{Brix1,Brix2}. Feng \textit{et al.} computed solutions to this equation using the theory of viscosity solutions \cite{Froese1,Froese2,Froese3} and Kawecki \textit{et al.} used a finite element method \cite{Kawecki}. Finally, Caboussat \textit{et al.} developed a least-squares method to compute solutions to the Monge-Ampère equation \cite{Caboussat}, which Prins \textit{et al.} adapted for transport boundary conditions \cite{Corien}. This method was extended by Yadav \textit{et al.} taking multiple freeform surfaces into account \cite{Nitin,Pieter}, and subsequently by Romijn \textit{et al.} taking generating functions as input \cite{Lotte,Jan2} and it is therefore also known as the \textit{generating least-squares method}.

The parallel-to-parallel reflector system converts a parallel emittance from a source $\mathcal{S}$ to a parallel irradiance at a target $\mathcal{T}$ using two reflectors. This system was studied by Yadav \textit{et al.} \cite{Nitin1} and Van Roosmalen \textit{et al.} \cite{Teun1,Teun}. To find the desired reflectors, they introduced source coordinates $\bm{x}$ and target coordinates $\bm{y}$, from which they derived the distance $u_1(\bm{x})$ from the source to the first reflector and the distance $u_2(\bm{y})$ from the target to the second reflector. The generating least-squares method is based on this formulation and uses a \textit{generating function} of the form
\begin{align}
    u_1(\bm{x})=G(\bm{x},\bm{y},u_2(\bm{y});V),
\end{align}
where $V$ is the optical path length between the source and the target. In the parallel-to-parallel reflector system, the optical path length $V$ is constant. 

In the parallel-to-two-target reflector system, the directions of rays vary at the first target and hence, $V$ is a function of the target coordinates $\bm{y}$. The generating function can then be generalized to
\begin{align}\label{eq:general_generating_function}
    u_1(\bm{x})=G(\bm{x},\bm{y},u_2(\bm{y});V(\bm{y})).
\end{align}
The reflectors can be used to control the spreading of light rays at the targets and create two different light distributions for two separate targets.

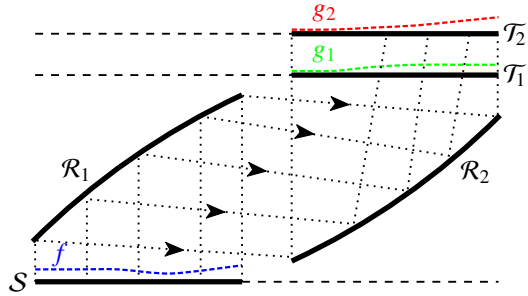
\begin{figure}[!ht]
\begin{center}
\begin{tikzpicture}[scale=0.55]
    \draw [line width=0.7mm] (0,0) -- (5,0);
    \draw [line width=0.7mm] (6.2,5) -- (11.2,5);
    \draw [line width=0.7mm] (6.2,6) -- (11.2,6);
    \draw [line width=0.7mm] (5,4.52) arc (115:135:17.8cm); 
    \draw [line width=0.7mm] (6.2,0.5) arc (295:315:17.7cm); 
    \draw [blue, dashed, line width=0.3mm, dash pattern=on 2pt off 1pt] plot[smooth] coordinates {(0,0.3) (1,0.3) (2,0.3) (3.3,0.2) (5,0.4)};
    \draw [green, dashed, line width=0.3mm, dash pattern=on 2pt off 1pt] plot[smooth] coordinates {(6.2,5.1) (7.2,5.1) (8.2,5.2) (9.2,5.25) (10.2,5.25) (11.2,5.25)};
    \draw [red, dashed, line width=0.3mm, dash pattern=on 2pt off 1pt] plot[smooth] coordinates {(6.2,6.1) (7.2,6.1) (8.2,6.15) (9.2,6.2) (10.2,6.3) (11.2,6.4)};
    \draw [dotted, line width=0.25mm] (0,0) -- (0,1) -- (6.2,0.6) -- (6.2,6); 
    \draw [dotted, line width=0.25mm] (1.25,0) -- (1.25,2) -- (7.7,1.4) -- (8.5,6); 
    \draw [dotted, line width=0.25mm] (2.5,0) -- (2.5,3.1) -- (9,2.2) -- (9.5,6); 
    \draw [dotted, line width=0.25mm] (4,0) -- (4,4) -- (10,3.05) -- (10.5,6); 
    \draw [dotted, line width=0.25mm] (5,0) -- (5,4.5) -- (11.18,4) -- (11.18,6); 
    \draw [arrows = {-Stealth[length=8pt, inset=2pt,round]}] (3.5,0.77) -- (3.6,0.765); 
    \draw [arrows = {-Stealth[length=8pt, inset=2pt,round]}] (4.5,1.69) -- (4.6,1.685); 
    \draw [arrows = {-Stealth[length=8pt, inset=2pt,round]}] (5.5,2.685) -- (5.6,2.675); 
    \draw [arrows = {-Stealth[length=8pt, inset=2pt,round]}] (6.6,3.591) -- (6.75,3.57); 
    \draw [arrows = {-Stealth[length=8pt, inset=2pt,round]}] (7.5,4.29) -- (7.6,4.284); 
    \draw [line width=0.25mm, dashed] (5,0) -- (11.2,0);
    \draw [line width=0.25mm, dashed] (0,5) -- (8,5);
    \draw [line width=0.25mm, dashed] (0,6) -- (8,6);
    \node at (0.6,0.65) [blue]{$f$};
    \node at (7,5.5) [green]{$g_1$};
    \node at (7,6.5) [red]{$g_2$};
    \node at (1,2.7) {$\mathcal{R}_1$};
    \node at (10.65,2.65) {$\mathcal{R}_2$};
    \node at (-0.4,0) {$\mathcal{S}$};
    \node at (11.6,5) {$\mathcal{T}_1$};
    \node at (11.6,6) {$\mathcal{T}_2$};
\end{tikzpicture}
\end{center}
\vspace{-0.5cm}
\caption{Parallel-to-two-target reflector system.}
\label{fig:two-reflector_two-target_formulation}
\end{figure}

In this paper we extend the generating least-squares method by modeling the two-target two-reflector system, illustrated in Fig.~\ref{fig:two-reflector_two-target_formulation}, where light rays originate from a parallel light source $\mathcal{S}$ with a specified emittance distribution $f$ indicated by the blue dashed line, then reflect via two reflectors $\mathcal{R}_1$ and $\mathcal{R}_2$, and finally reach the two targets $\mathcal{T}_1$ and $\mathcal{T}_2$. The goal is to find the shapes of the reflectors such that the light rays originating from $\mathcal{S}$ with emittance $f$ reach $\mathcal{T}_1$ and $\mathcal{T}_2$ with light distributions $g_1$ and $g_2$, respectively, indicated by the green and red dashed lines. As not all choices of $g_1$ and $g_2$ result in feasible reflectors, we will derive a \textit{feasibility condition} to determine when they are not self-intersecting. Notably, Feng \textit{et al.} also designed an optical system to control the spatial and directional target coordinates of light rays, but used a point source instead of a parallel source and lens surfaces instead of reflectors \cite{Feng2017}.


To find the shapes of the reflectors in the parallel-to-two-target reflector system, we apply a three-stage algorithm. In the first stage of the algorithm, we use energy conservation to determine the optical mapping from $\mathcal{T}_1$ to $\mathcal{T}_2$. In the second stage, we use this mapping to compute the optical path length $V$ from $\mathcal{S}$ to $\mathcal{T}_1$. In the final stage, we use the optical path length in a generated Jacobian equation and apply energy conservation to determine both the mapping from $\mathcal{S}$ to $\mathcal{T}_1$ and the shapes of the reflectors.

This paper is structured as follows. In Sec.~\ref{sec:Formulation} we model the parallel-to-two-target reflector system using generated Jacobian equations and the optical path length. In Sec.~\ref{sec:feasibility} we derive the feasibility condition and in Sec.~\ref{sec:algorithm} we provide an overview of the generating least-squares algorithm. We present results in Sec.~\ref{sec:results} and formulate our conclusions in Sec.~\ref{sec:conclusion}.

\section{Formulation of the parallel-to-two-target reflector system}\label{sec:Formulation}
In this section, we formulate a mathematical model for the parallel-to-two-target reflector system. We derive a generating function of the form (\ref{eq:general_generating_function}), find an equation for the optical path length and use energy balances to derive two generated Jacobian equations for the optical mappings. We use these results to compute the shapes of the reflectors.

We denote two-dimensional and three-dimensional vectors by a bold symbol, underlining them when they are in $\mathbb{R}^3$, and adding a hat to indicate unit length, e.g., $\underline{\hat{\bm{s}}}$. In addition, $|\cdot|$ denotes the 2-norm.

\subsection{The generating function formulation}
\begin{figure}[!ht]
\begin{center}
\begin{tikzpicture}[scale=1.0]
    \draw [line width=0.7mm] (0,0) -- (3,0); 
    \draw [line width=0.7mm] (3.2,3.5) -- (6.2,3.5); 
    \draw [line width=0.7mm] (3.2,4.5) -- (6.2,4.5); 
    \draw [line width=0.7mm] (3,2.9) arc (110:135:8.4cm); 
    \draw [line width=0.7mm] (3.2,0.68) arc (295:315:10.8cm); 
    \draw [line width=0.25mm, dashed] (-0.5,0) -- (6.6,0); 
    \draw [line width=0.25mm, dashed] (-0.5,3.5) -- (6.6,3.5); 
    \draw [line width=0.25mm, dashed] (-0.5,4.5) -- (6.6,4.5); 
    \draw [arrows = {-Stealth[length=8pt, inset=2pt,round]}] (-1,1.2) -- (-1,2.8); 
    \draw [arrows = {-Stealth[length=8pt, inset=2pt,round]}] (1.5965,0) -- (1.5965,1.5); 
    \draw [arrows = {-Stealth[length=8pt, inset=2pt,round]}] (1.5965,2.25) -- (3.1,1.8); 
    \draw [arrows = {-Stealth[length=8pt, inset=2pt,round]}] (4.5,1.4) -- (4.7585,3); 
    \draw [line width=0.25mm] (1.5965,1.4) -- (1.5965,2.25); 
    \draw [line width=0.25mm] (3,1.83) -- (4.5,1.4); 
    \draw [line width=0.25mm] (4.751,2.95) -- (5,4.5); 
    \node at (1.5965,0)[circle,fill,inner sep=1.8pt]{}; 
    \node at (1.5965,2.25)[circle,fill,inner sep=1.8pt]{}; 
    \node at (4.5,1.4)[circle,fill,inner sep=1.8pt]{}; 
    \node at (4.835,3.5)[circle,fill,inner sep=1.8pt]{}; 
    \node at (5,4.5)[circle,fill,inner sep=1.8pt]{}; 
    \node at (-1.3,2) {$z$};
    \node at (-1.3,0) {$z=0$};
    \node at (-1.2,3.5) {$z=L_1$};
    \node at (-1.2,4.5) {$z=L_2$};
    \node at (0,-0.3) {$\bm{0}$};
    \node at (1.25,0.3) {$\bm{x}$};
    \node at (1.4,2.55) {$P_1$};
    \node at (4.8,1.2) {$P_2$};
    \node at (4.585,3.8) {$\bm{y}$};
    \node at (4.75,4.8) {$\bm{z}$};
    \node at (1.8,0.8) {$\underline{\hat{\bm{s}}}$};
    \node at (2.2,1.75) {$\underline{\hat{\bm{i}}}$};
    \node at (4.9,2.3) {$\underline{\hat{\bm{t}}}$};
    \node at (0.2,1.8) {$\mathcal{R}_1$};
    \node at (5.8,1.8) {$\mathcal{R}_2$};
    \node at (2.2,-0.3) {$\mathcal{S}$};
    \node at (4,3.18) {$\mathcal{T}_1$};
    \node at (4,4.18) {$\mathcal{T}_2$};
\end{tikzpicture}
\end{center}
\caption{Parallel-to-two-target reflector system layout.}
\label{fig:2T_reflector_system}
\end{figure}
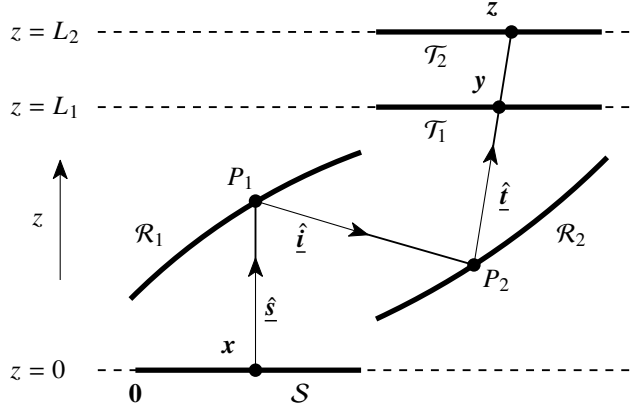

We consider the parallel-to-two-target reflector system illustrated in Fig.~\ref{fig:2T_reflector_system}, where a light ray originates from $(\bm{x},0)$ on the source plane $\mathcal{S}$ with direction $\underline{\hat{\bm{s}}}$ and hits the first reflector $\mathcal{R}_1$ at point $P_1$, with reflected direction $\underline{\hat{\bm{i}}}$. Thereafter, it hits the second reflector $\mathcal{R}_2$ at point $P_2$ and is reflected again with direction $\underline{\hat{\bm{t}}}$. The ray then reaches $(\bm{y},L_1)$ on the first target plane $\mathcal{T}_1$, maintains its direction $\underline{\hat{\bm{t}}}$ and finally reaches $(\bm{z},L_2)$ on the second target plane $\mathcal{T}_2$. In this optical system, $u_1=u_1(\bm{x})$ is the distance from $(\bm{x},0)$ to $P_1$, $d=d(\bm{x},\bm{y})$ is the distance from $P_1$ to $P_2$ and $u_2=u_2(\bm{y})$ is the distance from $P_2$ to $(\bm{y},L_1)$.

\noindent The reflectors can be parameterized as
\begin{align}\label{eq:parameterizations_reflectors}
    \mathcal{R}_1: \underline{\bm{r}}_1=\underline{\bm{r}}_1(\bm{x})=\begin{pmatrix}
        \bm{x}\\0
    \end{pmatrix}+u_1(\bm{x})\underline{\hat{\bm{s}}},
    &&
    \mathcal{R}_2: \underline{\bm{r}}_2=\underline{\bm{r}}_2(\bm{y},\bm{z})=\begin{pmatrix}
        \bm{y}\\L_1
    \end{pmatrix}-u_2(\bm{y})\underline{\hat{\bm{t}}},
\end{align}
where the source vector $\underline{\hat{\bm{s}}}=\underline{\hat{\bm{s}}}(\bm{x})$ and target vector $\underline{\hat{\bm{t}}}=\underline{\hat{\bm{t}}}(\bm{y},\bm{z})$ are given by
\begin{align}\label{eq:vectors_s_and_t}
    \underline{\hat{\bm{s}}}=\begin{pmatrix}
        \bm{p}_\mathrm{s}\\
        s_3
    \end{pmatrix}=\begin{pmatrix}
        \bm{0}\\
        1
    \end{pmatrix},
    &&
    \underline{\hat{\bm{t}}}=\begin{pmatrix}
        \bm{p}_\mathrm{t}\\
        t_3
    \end{pmatrix}
    =
    \frac{1}{\sqrt{|\bm{z}-\bm{y}|^2+(L_2-L_1)^2}}
    \begin{pmatrix}
        \bm{z}-\bm{y}\\
        L_2-L_1
    \end{pmatrix},
\end{align}
where $\bm{p}_\mathrm{s}$ and $\bm{p}_\mathrm{t}$ are the optical \text{momentum} of the light rays at the source and at the first target, respectively.

Next, we derive a generating function $G$ for the optical system of the form (\ref{eq:general_generating_function}) and its corresponding inverse $H$. To this end, we consider the optical path length $V$ from $\mathcal{S}$ to $\mathcal{T}_1$, which is the distance from the source to the first target along the path of a light ray, and can therefore be written as
\begin{align}\label{eq:2T_OPL_V}
    V=V(\bm{x},\bm{y})=u_1(\bm{x})+d(\bm{x},\bm{y})+u_2(\bm{y}).
\end{align}
Here, the square of $d=d(\bm{x},\bm{y})$ is given by
\begin{align}
    d^2&=\Bigg|\;\begin{pmatrix}
        \bm{y}-\bm{x}\\L_1
    \end{pmatrix}-u_1\underline{\hat{\bm{s}}}-u_2\underline{\hat{\bm{t}}}\;\Bigg|^2\nonumber\\
    &=|\bm{y}-\bm{x}|^2+L_1^2+u_1^2+u_2^2-2u_1 L_1-2u_2\Big((\bm{y}-\bm{x})\bm{\cdot}\bm{p}_\mathrm{t}+L_1t_3\Big)+2u_1u_2t_3.\label{eq:d_squared}
\end{align}
If we rewrite Eq.~(\ref{eq:2T_OPL_V}) as $d=V-u_1-u_2$ and square it, we have
\begin{align*}
    d^2=V^2+u_1^2+u_2^2-2Vu_1-2V u_2+2u_1u_2.
\end{align*}
By substituting Eq.~(\ref{eq:d_squared}) and rewriting, we then find
\begin{align*}
    V^2&=|\bm{y}-\bm{x}|^2+L_1^2+2u_1(V-L_1)+2u_2\big(V-\bm{p}_\mathrm{t}\bm{\cdot}(\bm{y}-\bm{x})-t_3L_1\big)-2u_1u_2(1-t_3).
\end{align*}
From this equation we obtain
\begin{subequations}
\begin{align}
    u_1(\bm{x})=G(\bm{x},\bm{y},u_2(\bm{y});V(\bm{y})),\label{eq:2T_u1_is_generating_function_G}
\end{align}
where $G=(\bm{x},\bm{y},w;V)$ is the generating function given by
\begin{align}
    G=G(\bm{x},\bm{y},w;V)&=\frac{\tfrac{1}{2}(V^2-|\bm{y}-\bm{x}|^2-L_1^2)-w(V-\bm{p}_\mathrm{t}\bm{\cdot}(\bm{y}-\bm{x})-t_3L_1)}{V-L_1-w(1-t_3)}.\label{eq:2T_generating_function_G}
\end{align}
\end{subequations}
From Eq. (\ref{eq:2T_u1_is_generating_function_G}) we can derive
\begin{subequations}
\begin{align}
    u_2(\bm{y})=H(\bm{x},\bm{y},u_1(\bm{x});V(\bm{y})),\label{eq:2T_u2_is_generating_function_H}
\end{align}
where $H=H(\bm{x},\bm{y},w;V)$ is the inverse of $G$, given by
\begin{align}
    H=H(\bm{x},\bm{y},w;V)&=\frac{\tfrac{1}{2}(V^2-|\bm{y}-\bm{x}|^2-L_1^2)-w(V-L_1)}{V-\bm{p}_\mathrm{t}\bm{\cdot}(\bm{y}-\bm{x})-t_3L_1- w(1-t_3)}. \label{eq:2T_inverse_generating_function_H}
\end{align}
\end{subequations}
A necessary condition for the generating function $G$ to have an inverse $H$ is that either $\partial_w G<0$ or $\partial_w G>0$ for all $\bm{x},\bm{y}\in\mathbb{R}^2$ and $w\in\mathbb{R}$ \cite[p. 116]{Lotte}. Differentiation gives
\begin{align}\label{eq:G_w}
    \partial_w G=\frac{\partial}{\partial w}\frac{\kappa_1-\kappa_2 w}{\kappa_3-\kappa_4 w}=\frac{\kappa_1 \kappa_4-\kappa_2 \kappa_3}{(\kappa_3-\kappa_4 w)^2},
\end{align}
with $\kappa_1=\tfrac{1}{2}(V^2-|\bm{y}-\bm{x}|^2-L_1^2)>0$, $\kappa_2=V-\bm{p}_\mathrm{t}\bm{\cdot}(\bm{y}-\bm{x})-t_3L_1>0$, $\kappa_3=V-L_1>0$ and $\kappa_4=1-t_3\geq 0$. We can use Eq. (\ref{eq:2T_OPL_V}) to write $\kappa_1=u_1d(1-\underline{\hat{\bm{s}}}\bm{\cdot}\underline{\hat{\bm{i}}})+u_1u_2(1-\underline{\hat{\bm{s}}}\bm{\cdot}\underline{\hat{\bm{t}}})+du_2(1-\underline{\hat{\bm{i}}}\bm{\cdot}\underline{\hat{\bm{t}}})$and $\kappa_2=(1-\underline{\hat{\bm{s}}}\bm{\cdot}\underline{\hat{\bm{t}}})u_1+(1-\underline{\hat{\bm{i}}}\bm{\cdot}\underline{\hat{\bm{t}}})d$. The numerator of $\partial_w G$ in expression (\ref{eq:G_w}) can then be written as $\kappa_1 \kappa_4-\kappa_2 \kappa_3=-(1-i_3)(1-\underline{\hat{\bm{i}}}\bm{\cdot}\underline{\hat{\bm{t}}})d^2$. Since $i_3\neq1$, we have $\kappa_1 \kappa_4 - \kappa_2 \kappa_3<0$ and from expression (\ref{eq:G_w}) we therefore conclude that $\partial_w G<0$.

\subsection{Energy conservation}
We will now derive relations for energy conservation between $\mathcal{S}$ and $\mathcal{T}_1$, as well as between $\mathcal{T}_1$ and $\mathcal{T}_2$, by defining optical mappings $\bm{m}_1:\mathcal{S}\to\mathcal{T}_1$ and $\bm{m}_2:\mathcal{T}_1\to\mathcal{T}_2$, where $\bm{m}_1(\bm{x})=\bm{y}$ and $\bm{m}_2(\bm{y})=\bm{z}$. We assume that $\mathcal{S}$ emits light with emittance $f:\mathcal{S}\to(0,\infty)$, which is then sent to $\mathcal{T}_1$ with illuminance $g_1:\mathcal{T}_1\to(0,\infty)$ and thereafter to $\mathcal{T}_2$ with illuminance $g_2:\mathcal{T}_2\to(0,\infty)$. Energy conservation between $\mathcal{S}$ and $\mathcal{T}_1$ tells us that
\begin{align*}
    \iint_{\mathcal{A}}f(\bm{x})\;\mathrm{d}\bm{x}=\iint_{\bm{m}_1(\mathcal{A})}g_1(\bm{y})\;\mathrm{d}\bm{y}=\iint_{\mathcal{A}}g_1(\bm{m}_1(\bm{x}))\;|\text{det}(\text{D}\bm{m}_1(\bm{x}))|\;\mathrm{d}\bm{x},
\end{align*}
for arbitrary $\mathcal{A}\subset \mathcal{S}$, where $\text{D}\bm{m}_1(\bm{x})$ denotes the Jacobian matrix of $\bm{m}_1(\bm{x})$. Similarly, energy conservation between $\mathcal{T}_1$ and $\mathcal{T}_2$ gives
\begin{align*}
    \iint_{\mathcal{B}}g_1(\bm{y})\;\mathrm{d}\bm{y}=\iint_{\bm{m}_2(\mathcal{B})}g_2(\bm{z})\;\mathrm{d}\bm{z}=\iint_{\mathcal{B}}g_2(\bm{m}_2(\bm{y}))\;|\text{det}(\text{D}\bm{m}_2(\bm{y}))|\;\mathrm{d}\bm{y},
\end{align*}
for arbitrary $\mathcal{B}\subset \mathcal{T}_1$. Therefore, assuming $\text{det}(\text{D}\bm{m}_1(\bm{x}))>0$ and $\text{det}(\text{D}\bm{m}_2(\bm{y}))>0$, we obtain the generated Jacobian equations
\begin{subequations}
\label{eq:2T_Monge_Ampere_equations}
\begin{align}
\label{eq:2T_Monge_Ampere_equations_1}
    \text{det}(\text{D}\bm{m}_1(\bm{x}))&=\frac{f(\bm{x})}{g_1(\bm{m}_1(\bm{x}))}=:F_1(\bm{x},\bm{m}_1(\bm{x})),\\
    \text{det}(\text{D}\bm{m}_2(\bm{y}))&=\frac{g_1(\bm{y})}{g_2(\bm{m}_2(\bm{y}))}=:F_2(\bm{y},\bm{m}_2(\bm{y})),\label{eq:2T_Monge_Ampere_equations_2}
\end{align}
\end{subequations}
which we will use for computing $\bm{y}=\bm{m}_1(\bm{x})$ and $\bm{z}=\bm{m}_2(\bm{y})$. 

In addition to the energy balances, we impose transport boundary conditions, which ensure that the boundary of the source domain is mapped to the boundary of the first target domain, and similarly, that the boundary of the first target domain is mapped to the boundary of the second target domain, i.e.,
\begin{subequations}
\begin{align}\label{eq:transport_boundary_condition_BC_1}
    \bm{m}_1(\partial \mathcal{S})&=\partial \mathcal{T}_1,\\
    \bm{m}_2(\partial \mathcal{T}_1)&=\partial \mathcal{T}_2.\label{eq:transport_boundary_condition_BC_2}
\end{align}
\end{subequations}
These conditions guarantee that all the light from the source will reach the second target via the first target \cite[p. 65-66]{Teun1}.

\subsection{Optical path length}\label{sec:Hamilton}

Under the assumption that optical mapping $\bm{z}=\bm{m}_2(\bm{y})$ is known, we compute the optical path length $V=V(\bm{x},\bm{y})$ using the equations
\begin{align}\label{eq:relation_V}
    \nabla_{\bm{x}}V=\frac{\partial V}{\partial \bm{x}}=-\bm{p}_\mathrm{s}=\bm{0},&&\nabla_{\bm{y}} V=\frac{\partial V}{\partial \bm{y}}=\bm{p}_t=\frac{\bm{z}-\bm{y}}{\sqrt{|\bm{z}-\bm{y}|^2+(L_2-L_1)^2}}.
\end{align}
A derivation of these equations can be found in \cite{Lotte,BornAndWolf}. The first equation states that the optical path length is independent of $\bm{x}$ and we can use the second equation to solve for $V=V(\bm{y})$.

\subsection{G-convexity}\label{sec:convexity_crit}
The generating functions $G(\bm{x},\bm{y},u_2(\bm{y});V(\bm{y}))$ and $H(\bm{x},\bm{y},u_1(\bm{x});V (\bm{y}))$ given by the expressions (\ref{eq:2T_generating_function_G}) and (\ref{eq:2T_inverse_generating_function_H}) are each others inverses. As a result, there are many pairs $(u_1(\bm{x}),u_2(\bm{y}))$ satisfying the corresponding Eqs. (\ref{eq:2T_u1_is_generating_function_G}) and (\ref{eq:2T_u2_is_generating_function_H}). We select a unique solution by enforcing $u_1$ and $u_2$ to form either a $G-$convex or $G-$concave pair, see \cite[p. 239]{Lotte}. In the first case, the solution is given by
\begin{align*}
    u_1(\bm{x})&=\max_{\bm{y}\in\mathcal{T}_1}G(\bm{x},\bm{y},u_2(\bm{y});V(\bm{y})),&&u_2(\bm{y})=\max_{\bm{x}\in\mathcal{S}}H(\bm{x},\bm{y},u_1(\bm{x});V(\bm{y})),
\end{align*}
and in the second case, the solution reads
\begin{align*}
    u_1(\bm{x})&=\min_{\bm{y}\in\mathcal{T}_1}G(\bm{x},\bm{y},u_2(\bm{y});V(\bm{y})),&&u_2(\bm{y})=\min_{\bm{x}\in\mathcal{S}}H(\bm{x},\bm{y},u_1(\bm{x});V(\bm{y})).
\end{align*}
Since in both cases $\widetilde{H}(\bm{x},\bm{y}):=H(\bm{x},\bm{y},u_1(\bm{x});V(\bm{y}))$ has a stationary point with respect to $\bm{x}$, we obtain the necessary condition
\begin{align}\label{eq:grad_H_tilde}
    \nabla_{\bm{x}}\widetilde{H}(\bm{x},\bm{y})=\nabla_{\bm{x}}H(\bm{x},\bm{y},u_1(\bm{x});V(\bm{y}))+\partial_w H(\bm{x},\bm{y},u_1(\bm{x});V(\bm{y}))\nabla u_1(\bm{x})=\bm{0}.
\end{align}
By the implicit function theorem, this equation provides a mapping $\bm{y}=\bm{m}_1(\bm{x})$ under the condition that the mixed Hessian matrix
\begin{align}\label{eq:matrix_C}
    \bm{C}=\bm{C}(\bm{x},\bm{y},u_1(\bm{x});V(\bm{y}))=\text{D}_{\bm{xy}}\widetilde{H}(\bm{x},\bm{y})=\left(\frac{\partial^2\widetilde{H}(\bm{x},\bm{y})}{\partial x_i\,\partial y_j}\right),
\end{align}
is invertible for all $\bm{x}\in\mathcal{S}$ and $\bm{y}\in\mathcal{T}_1$. For our simulations, we verified numerically that this holds for Eq.~(\ref{eq:2T_inverse_generating_function_H}). After substituting $\bm{y}=\bm{m}_1(\bm{x})$ into Eq.~(\ref{eq:grad_H_tilde}) and subsequently taking the derivative with respect to $\bm{x}$, we find
\begin{align}\label{eq:CDmP}
    \bm{C}\text{D}\bm{m}_1=-\text{D}_{\bm{xx}}\widetilde{H}(\bm{x},\bm{y})=:\bm{P}(\bm{x},\bm{y},u_1(\bm{x});V(\bm{y}))=\bm{P},
\end{align}
where $\text{D}_{\bm{xx}}\widetilde{H}$ denotes the Hessian matrix of $\widetilde{H}$ with respect to $\bm{x}$. For a $G-$convex pair $(u_1(\bm{x}),u_2(\bm{y}))$, the matrix $\bm{P}$ must be symmetric positive definite (SPD). Similarly, for a $G-$concave pair $(u_1(\bm{x}),u_2(\bm{y}))$, the matrix $\bm{P}$ must be symmetric negative definite (SND). In addition, from Eq.~(\ref{eq:2T_Monge_Ampere_equations_1}) we also require $\bm{P}$ to satisfy
\begin{align}\label{eq:detP1_constraint}
    \text{det}(\bm{P})=F_1(\cdot,\bm{m}_1)\text{det}(\bm{C}).
\end{align}

\subsection{Summary}\label{sec:summary}
In this section, we presented a model for the parallel-to-two-target two-reflector system. We will find the shapes of the reflectors as follows. 
\begin{itemize}
    \item We compute the optical mapping $\bm{z}=\bm{m}_2(\bm{y})$ from Eq.~(\ref{eq:2T_Monge_Ampere_equations_2}) subject to Eq.~(\ref{eq:transport_boundary_condition_BC_2}). This procedure will be explained in Sec.~\ref{sec:algorithm}.
    \item We find the optical path length $V(\bm{y})$ from Eq.~(\ref{eq:relation_V}).
    \item We solve Eq.~(\ref{eq:CDmP}) for $\bm{y}=m_1(\bm{x})$ subject to the constraint (\ref{eq:detP1_constraint}) and the transport boundary condition (\ref{eq:transport_boundary_condition_BC_1}).
    \item We solve Eq.~(\ref{eq:grad_H_tilde}) for $u_1$, substitute this expression in (\ref{eq:2T_u2_is_generating_function_H}) to find $u_2$ and finally use the reflector parameterizations given in Eqs. (\ref{eq:parameterizations_reflectors}) to find the shapes of the reflectors.
\end{itemize}

\section{Feasibility condition}\label{sec:feasibility}
In the two-target two-reflector system, the desired illuminances $g_1$ and $g_2$ on the two target planes will in general not guarantee a physically working reflector system. As an example, after the computation of the reflector surfaces, the shape of the second reflector might self-intersect, causing a reflector that is impossible to manufacture. In this case, we call the reflector \textit{self-intersecting}. An example is shown in Fig. \ref{fig:2D_system3}. To prevent the reflector from self-intersecting, we will derive a feasibility condition.

\begin{figure}[!ht]
  \centering
\begin{subfigure}[b]{0.43\textwidth}
    \centering
    \begin{tikzpicture}[>=latex,scale=1.5]
    \fill[color=gray!50!white] (1.9,0.1) -- (3,1.6) -- (3,3.5) -- (1.9,2.4) -- cycle;
    \draw[arrows = {-Stealth[length=8pt, inset=2pt,round]}] (3,1.6) -- (2.2,0.5); 
    \draw[arrows = {-Stealth[length=8pt, inset=2pt,round]}] (3,1.6) -- (4.35,1.38); 
    \draw[arrows = {-Stealth[length=8pt, inset=2pt,round]}] (3,1.6) -- (3,2.8); 
    
    \draw[arrows = {-Stealth[length=8pt, inset=2pt,round]}] (3,1.6) -- (2.7,2.2); 
    \draw[arrows = {-Stealth[length=8pt, inset=2pt,round]}] (3,1.6) -- (4,2); 
    \draw[line width=0.25mm, dotted] (4,2) -- (2.7,2.2); 
    
    \draw[arrows = {-Stealth[length=8pt, inset=2pt,round]}] (3,1.6) -- (2.3,1.1); 
    \draw[arrows = {-Stealth[length=8pt, inset=2pt,round]}] (3,1.6) -- (2.8,1); 
    \draw[line width=0.25mm, dotted] (2.8,1) -- (2.3,1.1); 

    \coordinate (d) at (2.8,2);
    \coordinate (e) at (3,1.6);
    \coordinate (f) at (2.6,1.3);
    \pic [draw, angle radius=0.4cm, angle eccentricity=1.5] {angle = d--e--f};

    \node at (2.1,0.8) {$\underline{\hat{\bm{e}}}_1$}; 
    \node at (4.25,1.15) {$\underline{\hat{\bm{e}}}_2$}; 
    \node at (3.25,2.7) {$\underline{\hat{\bm{e}}}_z$}; 
    \node at (3.1,1.4) {$\bm{r}$}; 
    \node at (4.2,2) {$\underline{\hat{\bm{t}}}_1$}; 
    \node at (3,0.9) {$\underline{\hat{\bm{t}}}_2$}; 
    \node at (2.55,1.7) {$\theta_1$}; 
    \node at (2.6,2.4) {$\underline{\hat{\bm{t}}}_{1,p}$}; 
    \node at (2.1,1.2) {$\underline{\hat{\bm{t}}}_{2,p}$}; 
    \node at (3,1.6)[circle,fill,inner sep=1.8pt]{}; 
    \end{tikzpicture}
    \caption{Directions projected on the $(\underline{\hat{\bm{e}}}_1,\hat{\underline{\bm{e}}}_z)$-plane.}
    \label{fig:difference_zenith_angle}
\end{subfigure}
\hfill
\begin{subfigure}[b]{0.43\textwidth}
    \begin{center}
\begin{tikzpicture}[scale=1.0]
    \draw [line width=0.7mm] (3.2,3.5) -- (6.2,3.5); 
    \draw [line width=0.7mm] (3.2,4.5) -- (6.2,4.5); 
    \draw [line width=0.7mm] (3.2,0.68) arc (295:310:10.8cm); 
    \draw [line width=0.25mm, dashed] (2.3,3.5) -- (6.6,3.5); 
    \draw [line width=0.25mm, dashed] (2.3,4.5) -- (6.6,4.5); 
    \draw [arrows = {-Stealth[length=8pt, inset=2pt,round]}] (7,0.85) -- (7,2.25); 
    \draw [arrows = {-Stealth[length=8pt, inset=2pt,round]}] (7,0.85) -- (8.4,0.85); 
    \draw [arrows = {-Stealth[length=8pt, inset=2pt,round]}] (3.8,1) -- (3.89,2.5); 
    \draw [arrows = {-Stealth[length=8pt, inset=2pt,round]}] (3.8,1) -- (4.48,2.4); 
    \draw [line width=0.25mm] (3.88,2.4) -- (4,4.5); 
    \draw [line width=0.25mm] (4.45,2.33) -- (5.5,4.5); 
    \draw [line width=0.25mm] (4.82,3.1) -- (3.95,3.5); 
    \draw [line width=0.25mm, dotted] (4.82,3.1) -- (5.68,2.7); 
    \draw [line width=0.25mm, dotted] (2.3,2.2) -- (3.8,1); 
    \draw [line width=0.25mm, dotted] (2.3,1.3) -- (3.8,1); 
    \node at (3.8,1)[circle,fill,inner sep=1.8pt]{}; 
    \node at (3.95,3.5)[circle,fill,inner sep=1.8pt]{}; 
    \node at (5.03,3.5)[circle,fill,inner sep=1.8pt]{}; 
    \node at (4,4.5)[circle,fill,inner sep=1.8pt]{}; 
    \node at (5.5,4.5)[circle,fill,inner sep=1.8pt]{}; 
    \node at (4.82,3.1)[circle,fill,inner sep=1.8pt]{}; 
    \coordinate (y1) at (3.95,3.5);
    \coordinate (y2) at (5.03,3.5);
    \coordinate (z1) at (4,4.5);
    \coordinate (z2) at (5.5,4.5);
    \coordinate (r) at (3.8,1);
    \coordinate (w) at (4.82,3.1);
    \pic [draw, angle radius=0.8cm, angle eccentricity=1.5] {angle = z2--r--z1};
    \pic [draw, angle radius=0.3cm, angle eccentricity=1.5] {angle = r--y1--w};
    \pic [draw, angle radius=0.45cm, angle eccentricity=1.5] {angle = y2--y1--z1};
    \pic [draw, angle radius=1.9cm, angle eccentricity=1.5] {angle = w--y1--y2};
    \draw [line width=0.15mm] (4.7,2.85) -- (4.45,2.97) -- (4.565,3.2);
    \node at (4.05,2) {$\theta_1$};
    \node at (4.15,3) {$\theta_2$};
    \node at (6.05,3.05) {$\theta_3$};
    \node at (4.4,4) {$\theta_4$};
    \node at (5.25,3.2) {$\Delta u_2$};
    \node at (6.7,1.55) {$\underline{\hat{\bm{e}}}_z$};
    \node at (7.7,0.55) {$\underline{\hat{\bm{e}}}_1$};
    \node at (3.95,0.75) {$\bm{r}$};
    \node at (4.9,2.8) {$\bm{w}$};
    \node at (3.75,3.75) {$\bm{y}_1$};
    \node at (5.45,3.75) {$\bm{y}_2$};
    \node at (4,4.75) {$\bm{z}_1$};
    \node at (5.5,4.75) {$\bm{z}_2$};
    \node at (3.5,2.2) {$\underline{\hat{\bm{t}}}_{1,p}$};
    \node at (4.85,2.2) {$\underline{\hat{\bm{t}}}_{2,p}$};
    \node at (5.8,1.8) {$\mathcal{R}_2$};
    \node at (3.2,3.18) {$\mathcal{T}_1$};
    \node at (3.2,4.18) {$\mathcal{T}_2$};
\end{tikzpicture}
\end{center}
\caption{Rays projected on the $(\underline{\hat{\bm{e}}}_1,\underline{\hat{\bm{e}}}_z)-$plane.}
\label{fig:2T_reflector_system_overlap}
\end{subfigure}
\caption{Two rays reflecting in the same point $\bm{r}$ on the second reflector $\mathcal{R}_2$.}
\end{figure}
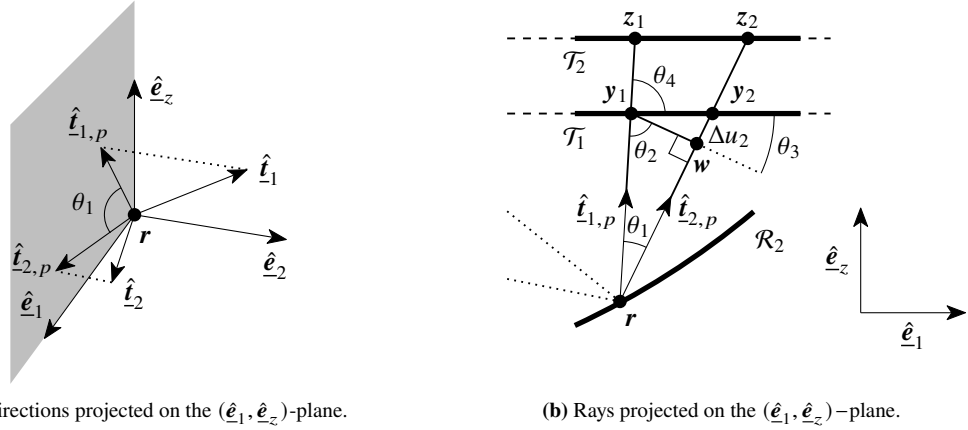

To this end, we focus on the projection of two distinct rays with directions $\underline{\hat{\bm{t}}}_1$ and $\underline{\hat{\bm{t}}}_2$ on the $(\underline{\hat{\bm{e}}}_1,\underline{\hat{\bm{e}}}_z)$-plane that both hit the second reflector $\mathcal{R}_2$ in the same point $\bm{r}$ and would hereby violate uniqueness of the reflector. We denote their projection vectors by $\underline{\hat{\bm{t}}}_{1,p}$ and $\underline{\hat{\bm{t}}}_{2,p}$ and define $\theta_1$ as their zenith angle difference, as illustrated in Fig.~\ref{fig:difference_zenith_angle}. We let $\bm{y}_1,\bm{y}_2$ and $\bm{z}_1,\bm{z}_2$ be the points on this plane at which the rays hit the first and second target plane, respectively, as illustrated in Fig.~\ref{fig:2T_reflector_system_overlap}. Without loss of generality, we assume that the length of the second ray from $\mathcal{R}_2$ to $\mathcal{T}_1$ is larger than the first ray and define $\Delta u_2$ to be their difference. Then, by $\bm{w}$ we denote the point on the second ray so that the line segment between $\bm{w}$ and $\bm{y}_1$ is perpendicular to the line segment between $\bm{r}$ and $\bm{y}_2$. Finally, we let $\Delta y_1$ be the distance between $\bm{y}_1$ and $\bm{y}_2$, and define the angles $\theta_2=\angle \bm{r}\bm{y}_1\bm{w}$, $\theta_3=\angle \bm{w}\bm{y}_1\bm{y}_2$ and $\theta_4=\angle \bm{y}_2\bm{y}_1\bm{z}_1$.

We can write $\sin(\theta_3)=\Delta u_2/\Delta y_1$. As $\underline{\hat{\bm{t}}}_{2,p}\to\underline{\hat{\bm{t}}}_{1,p}$, we have $\theta_1\to0$ and therefore $\theta_2\to\tfrac{\pi}{2}$. Moreover, since $\theta_2+\theta_3+\theta_4=\pi$, this implies that $\theta_3\to\tfrac{\pi}{2}-\theta_4$. Combining these observations, we find for $\theta_1\to0$
\begin{align*}
    \frac{\partial u_2}{\partial y_1}=\sin(\theta_3)=\cos{\left(\theta_4\right)}=\hat{\bm{e}}_1\bm{\cdot}\underline{\hat{\bm{t}}}_{1,p}=\frac{\partial V}{\partial y_1},
\end{align*}
where the last equation follows from the second relation of Eq.~(\ref{eq:relation_V}). Similarly, for the orthogonal projection on the $(\underline{\hat{\bm{e}}}_2,\underline{\hat{\bm{e}}}_z)$-plane we find $\frac{\partial u_2}{\partial y_{2}}=\frac{\partial V}{\partial y_{2}}$. From this analysis we conclude that $\mathcal{R}_2$ self-intersects when $\nabla_{\bm{y}} u_2=\nabla_{\bm{y}}V$. Thus, the second reflector does not self-intersect when $\nabla_{\bm{y}}\left(u_2-V\right)\neq \bm{0}$. Since two distinct rays originating from the source are parallel, they cannot hit $\mathcal{R}_1$ in the same point and the first reflector can therefore not self-intersect.

\section{The least-squares algorithm}\label{sec:algorithm}
To solve the two-target two-reflector system, we apply a three-stage least-squares algorithm. 

\noindent \textbf{Stage 1:} The first stage of the algorithm computes the mapping $\bm{z}=\bm{m}_2(\bm{y})$. Since this computation is a simpler version of the computation of $\bm{y}=\bm{m}_1(\bm{x})$, we first elaborate the procedure to compute $\bm{y}=\bm{m}_1(\bm{x})$ (see Stage~3) and outline the computation of $\bm{z}=\bm{m}_2(\bm{y})$ at the end of this section.

\noindent \textbf{Stage 2:} The second stage of the algorithm proceeds by computing the optical path length $V$. In the algorithm, the second relation of Eq.~(\ref{eq:relation_V}) is imposed in a least-squares sense by minimizing the functional
\begin{align}\label{eq:functional_L}
    K[\bm{m}_2,V]=\tfrac{1}{2}\iint_{\mathcal{T}_1}|\nabla V-\bm{p}_\mathrm{t}(\bm{y},\bm{m}_2(\bm{y}))|^2\;\mathrm{d}\bm{y}.
\end{align}
Specifically, the algorithm finds $V\in C^2(\mathcal{T}_1)$ from
\begin{align}
    V=\underset{\psi\in C^2(\mathcal{T}_1)}{\mathrm{argmin}}\;K[\bm{m}_2,\psi],
\end{align}
by solving the corresponding Neumann problem
\begin{subequations}
\label{eq:BVP_for_V1}
\begin{align}
    \nabla^2 V&=\nabla \bm{\cdot} \bm{p}_\mathrm{t}, && \bm{y}\in\mathcal{T}_1,\\
    \nabla V\bm{\cdot}\hat{\bm{n}}&=\bm{p}_\mathrm{t}\bm{\cdot}\hat{\bm{n}}, && \bm{y}\in\partial \mathcal{T}_1,
\end{align}
\end{subequations}
where $\hat{\bm{n}}$ is the unit outward normal on $\partial \mathcal{T}_1$. We derived this boundary value problem by setting the first variation of $K$ with respect to $V$ to zero, and solved it using a cell-centered finite volume method on a Cartesian grid. Since the solution of this boundary value problem is determined up to an additive constant, we enforce a unique solution by setting the value of $V$ for a specific light ray, such as the center ray, equal to some value $V_0$.

\noindent \textbf{Stage 3:} The final stage of the algorithm iteratively computes the mapping $\bm{y}=\bm{m}_1(\bm{x})$ and the reflector surfaces. To compute $\bm{y}=\bm{m}_1(\bm{x})$, the algorithm starts from an initial guess $\bm{m}_1^0$, which maps the smallest bounding box enclosing source domain $\mathcal{S}$ to the smallest bounding box enclosing target domain $\mathcal{T}_1$. Subsequently, the algorithm minimizes the functionals
\begin{subequations}
\begin{align}
    J_{I}[\bm{m}_1,\bm{P}]&=\tfrac{1}{2}\iint_{\mathcal{S}}||\bm{C}\text{D}\bm{m}_1-\bm{P}||_F^2\;\mathrm{d}\bm{x},\label{eq:functional_J_I}
    \\
    J_\mathrm{B}[\bm{m}_1,\bm{b}]&=\tfrac{1}{2}\oint_{\partial \mathcal{S}}|\bm{m}_1-\bm{b}|^2\;\mathrm{d}s,\label{eq:functional_J_B}
    \\
    J[\bm{m}_1,\bm{P},\bm{b}]&=\alpha J_\mathrm{I}[\bm{m}_1,\bm{P}]+(1-\alpha)J_\mathrm{B}[\bm{m}_1,\bm{b}],\label{eq:functional_J}
    \\
    I[\bm{m}_1,u_1]&=\tfrac{1}{2}\iint_{\mathcal{S}}\Big|\frac{\nabla_{\bm{x}} H(\bm{x},\bm{m}_1,u_1;V)}{\partial_w H(\bm{x},\bm{m}_1,u_1;V)}+\nabla u_1\Big|^2 \mathrm{d}\bm{x},\label{eq:functional_I}
\end{align}
with $||\cdot||_F$ the Frobenius norm, by computing
\begin{align}
    \bm{P}^{n+1}&=\underset{\bm{P}\in\mathcal{P}}{\mathrm{argmin}}\;J_\mathrm{I}[\bm{m}_1^n,\bm{P}],\label{eq:computing_1}
    \\
    \bm{b}^{n+1}&=\underset{\bm{b}\in\mathcal{B}}{\mathrm{argmin}}\;J_\mathrm{B}[\bm{m}_1^n,\bm{b}],\label{eq:computing_2}
    \\
    \bm{m}_1^{n+1}&=\underset{\bm{m}_1\in C^2(\mathcal{S})^2}{\mathrm{argmin}}\;J[\bm{m}_1,\bm{P}^{n+1},\bm{b}^{n+1}],\label{eq:computing_3}
    \\
    u_1^{n+1}&=\underset{u_1\in C^2(\mathcal{S})}{\mathrm{argmin}}\;I[\bm{m}_1^{n+1},u_1],\label{eq:computing_4}
\end{align}
where the function spaces $\mathcal{P}$ and $\mathcal{B}$ are given by
\begin{align}\label{eq:mathcalP}
    \mathcal{P}&=\{\bm{P}\in C^1(\mathcal{S})^{2\times2}\;\big|\; \bm{P}\text{ is SPD or SND, }\text{det}(\bm{P})=F_1(\bm{x},\bm{m}_1(\bm{x}))\text{det}(\bm{C})\},
    \\
    \mathcal{B}&=\{\bm{b}\in C^1(\partial \mathcal{S})^2\;\big|\;\bm{b}(\bm{x})\in\partial\mathcal{T}_1\}.
\end{align}
\label{eq:iteration_scheme}
\end{subequations}
In the first two steps of Stage 3, the functionals $J_\mathrm{I}$ and $J_\mathrm{B}$ are minimized for $\bm{P}$ and $\bm{b}$ to satisfy Eq.~(\ref{eq:CDmP}) subject to the constraint (\ref{eq:detP1_constraint}) and the transport boundary condition (\ref{eq:transport_boundary_condition_BC_1}). Here, both the mapping $\bm{z}=\bm{m}_2(\bm{y})=\bm{m}_2(\bm{m}_1(\bm{x}))$ and the optical path length $V(\bm{y})$ required to compute the matrices $\bm{C}$ and $\bm{P}$ are calculated using bilinear interpolation. The third step then minimizes a linear combination $J$ with weighting factor $\alpha\in(0,1)$ of both functionals $J_\mathrm{I}$ and $J_\mathrm{B}$ to update the mapping $\bm{y}=\bm{m}_1(\bm{x})$. The fourth step of the scheme minimizes the functional $I$ to satisfy Eq.~(\ref{eq:grad_H_tilde}). At the end of each iteration, we update $H(\bm{x},\bm{y},u_1(\bm{x});V(\bm{y}))$ and the matrices $\bm{C}$ and $\bm{P}$ using Eqs. (\ref{eq:2T_inverse_generating_function_H}), (\ref{eq:matrix_C}) and (\ref{eq:CDmP}). These steps are then repeated for $n=0,1,2,...$ until convergence. Finally, we use Eqs. (\ref{eq:parameterizations_reflectors}) to compute the shapes of the reflectors.

An in-depth analysis of the minimization of the functionals $J_\mathrm{I}$, $J_\mathrm{B}$ and $J$ that uses finite volume discretizations can be found in Romijn \textit{et al.} \cite{Lotte}. By writing the functional $I$ as
\begin{align}\label{eq:functional_I_new}
    I[u_1,\bm{m}_1]=\tfrac{1}{2}\iint_\mathcal{S}\left|\bm{f}+\nabla u_1\right|^2\mathrm{d}\bm{x}, &&
    \bm{f}=\bm{f}(\bm{x},\bm{m}_1,u_1;V)=\frac{\nabla_{\bm{x}}H(\bm{x},\bm{m}_1,u_1;V)}{\partial_w H(\bm{x},\bm{m}_1,u_1;V)},
\end{align}
and setting its first variation with respect to $u_1$ to zero, we obtain the boundary value problem
\begin{subequations}
\label{eq:BVP}
\begin{align}
    \nabla \bm{\cdot} (\bm{f}+\nabla u_1)&=\partial_w\bm{f}\bm{\cdot}(\bm{f}+\nabla u_1), && \bm{x}\in\mathcal{S},\\
    (\bm{f}+\nabla u_1)\bm{\cdot}\hat{\bm{n}}&=0, && \bm{x}\in\partial \mathcal{S},
\end{align}
\end{subequations}
where $\hat{\bm{n}}$ is the unit outward normal on $\partial \mathcal{S}$. To solve this boundary value problem for $u_1$, we apply a cell-centered finite volume method on a Cartesian grid. We verified that the compatibility condition $\iint_\mathcal{S} \partial_w \bm{f}\bm{\cdot}(\bm{f}+\nabla u_1)\mathrm{d}\bm{x}=0$ holds in each of our simulations. The solution of boundary value problem (\ref{eq:BVP}) is determined up to an additive constant. Therefore, we enforce a unique solution by choosing the value of $u_1$ for a specific light ray, such as the center ray, equal to a value $u_1(\bm{x}_c)$.

\noindent \textbf{Stage 1 revisited:} We compute $\bm{z}=\bm{m}_2(\bm{y})$ in a similar way as $\bm{y}=\bm{m}_1(\bm{x})$. However, since there are no optical surfaces between target planes $\mathcal{T}_1$ and $\mathcal{T}_2$, we can choose the generating function
\begin{align}\label{eq:generating_function_G_target}
    G_\mathcal{T}=G_\mathcal{T}(\bm{x},\bm{y},w)=\bm{x}\bm{\cdot}\bm{y}+w,
\end{align}
and its inverse
\begin{align}\label{eq:generating_function_H_target}
    H_\mathcal{T}=H_\mathcal{T}(\bm{x},\bm{y},w)=-\bm{x}\bm{\cdot}\bm{y}+w,
\end{align}
corresponding to a quadratic cost function, for which a derivation can be found in \mbox{\cite[p. 71]{Lotte}}. To compute $\bm{z}=\bm{m}_2(\bm{y})$, we use the iteration scheme~(\ref{eq:iteration_scheme}) once more subject to the following modifications:
\begin{itemize}
    \item In each of the expressions in iteration scheme (\ref{eq:iteration_scheme}), $\bm{m}_1$ is replaced by $\bm{m}_2$, $\mathcal{T}_1$ is replaced by $\mathcal{T}_2$, both $\mathcal{S}$ and $\partial\mathcal{S}$ are replaced by $\mathcal{T}_1$ and $\partial\mathcal{T}_1$, $H(\bm{x},\bm{m}_1,u_1;V)$ is replaced by $H_\mathcal{T}(\bm{x},\bm{y},w)$, and $F_1$ is replaced by $F_2$.
    \item The function $\widetilde{H}(\bm{x},\bm{y})$, used in expressions (\ref{eq:matrix_C}) and (\ref{eq:CDmP}) for matrices $\bm{C}$ and $\bm{P}$, is defined in terms of generating function $H_\mathcal{T}$ instead of $H$.
\end{itemize}

\section{Numerical results}\label{sec:results}
In this section we provide several examples of the two-target reflector system. We will first consider the two-dimensional model and provide two examples: a standard example with different light distributions and an example which verifies the feasibility condition discussed in Sec.~\ref{sec:feasibility}. We then consider the three-dimensional model and provide two examples that demonstrate the potential of the least-squares algorithm.

\subsection{Two-dimensional results}
In the two-dimensional model, $\mathcal{S}$, $\mathcal{T}_1$ and $\mathcal{T}_2$ are line segments, and $\hat{\bm{s}},\hat{\bm{t}}\in\mathbb{R}^2$. As a consequence, Eqs. (\ref{eq:2T_Monge_Ampere_equations}) reduce to ordinary differential equations (ODEs) that can be solved for the mappings $y=m_1(x)$ and $z=m_2(y)$. Eq.~(\ref{eq:relation_V}) also simplifies to the ODE $V'=\tfrac{\partial V}{\partial y}=t_1$, which can be solved for $V=V(y)$. Finally, taking the partial derivative on both sides of Eq.~(\ref{eq:2T_u2_is_generating_function_H}) with respect to $x$ and rearranging terms gives the ODE
\begin{subequations}
\label{eq:2T_gradient_generating_function_G}
\begin{align}
    u_1'= \frac{a_1+a_2 u_1}{a_3},
\end{align}
with auxiliary variables
\begin{align}
    a_1 &=(V-(y-x)t_1-L_1 t_2)(y-x)-\tfrac{1}{2}(V^2-(y-x)^2-L_1^2)t_1,\\
    a_2 &=(V-L_1)t_1-(1-t_2)(y-x),\\
    a_3 &= (V-L_1)(V-(y-x,L_1)^\mathrm{T}\bm{\cdot}\hat{\bm{t}})-\tfrac{1}{2}(V^2-(y-x)^2-L_1^2)(1-t_2),
\end{align}
\end{subequations}
which can be solved for $u_1$. As initial conditions for the ODEs, we can specify $V$ and $u_1$ for a specific light ray. Finally, $u_2$ can be obtained from Eq.~(\ref{eq:2T_u2_is_generating_function_H}) and Eqs. (\ref{eq:parameterizations_reflectors}) can be used to find the reflectors. Thus, applying the least-squares algorithm for the two-dimensional model is unnecessary and we therefore simply solve the above ODEs using Matlab's solver \textit{ode45}.

\begin{figure}[!ht]
\centering
\begin{subfigure}[b]{0.48\textwidth}
    \centering
    \includegraphics[width=1.0\textwidth,trim={0cm 0cm 0cm 0cm},clip]{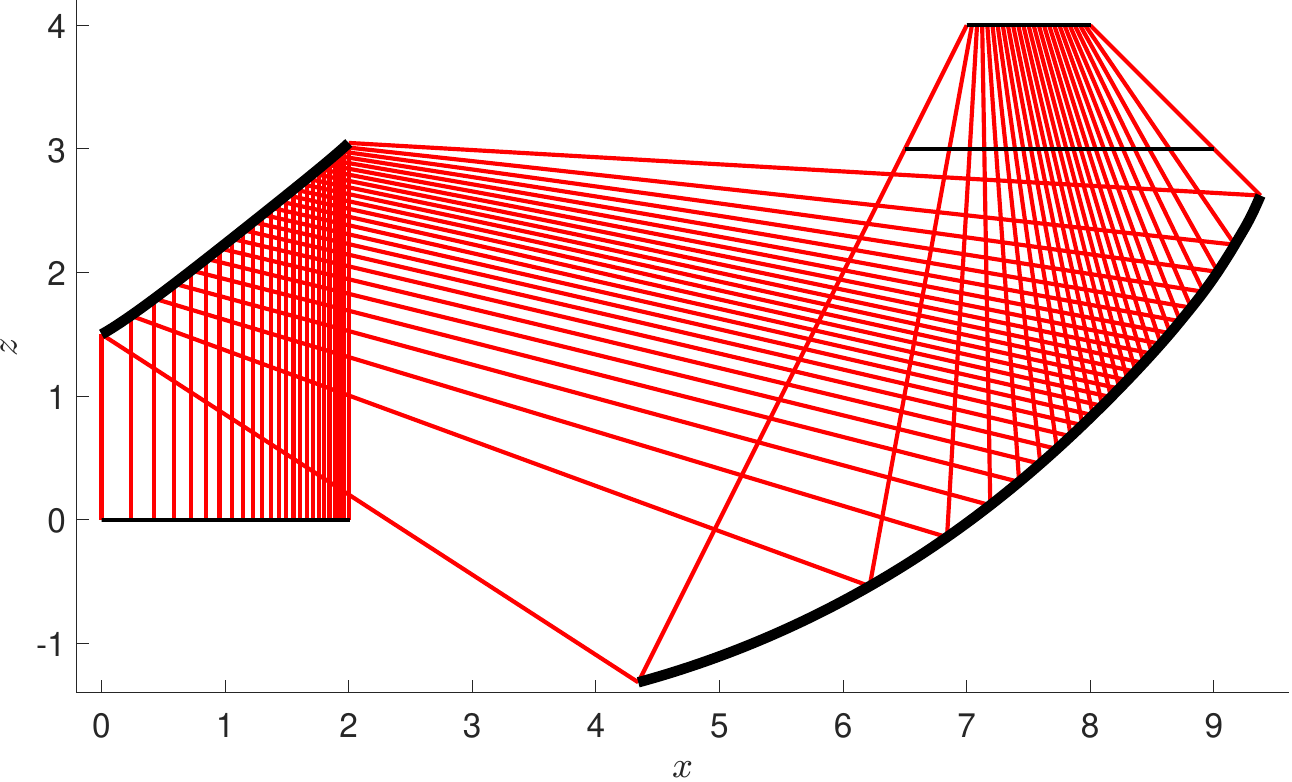}
    \caption{A standard example.}
    \label{fig:2D_system1}
\end{subfigure}
\begin{subfigure}[b]{0.48\textwidth}
    \begin{tikzpicture}[spy using outlines={circle,black,magnification=2.5,size=2.5cm, connect spies}]
    \node [anchor=south west,inner sep=0] (image) at (0,0) {\includegraphics[width=1.0\textwidth,trim={0cm 0cm 0cm 0cm},clip]{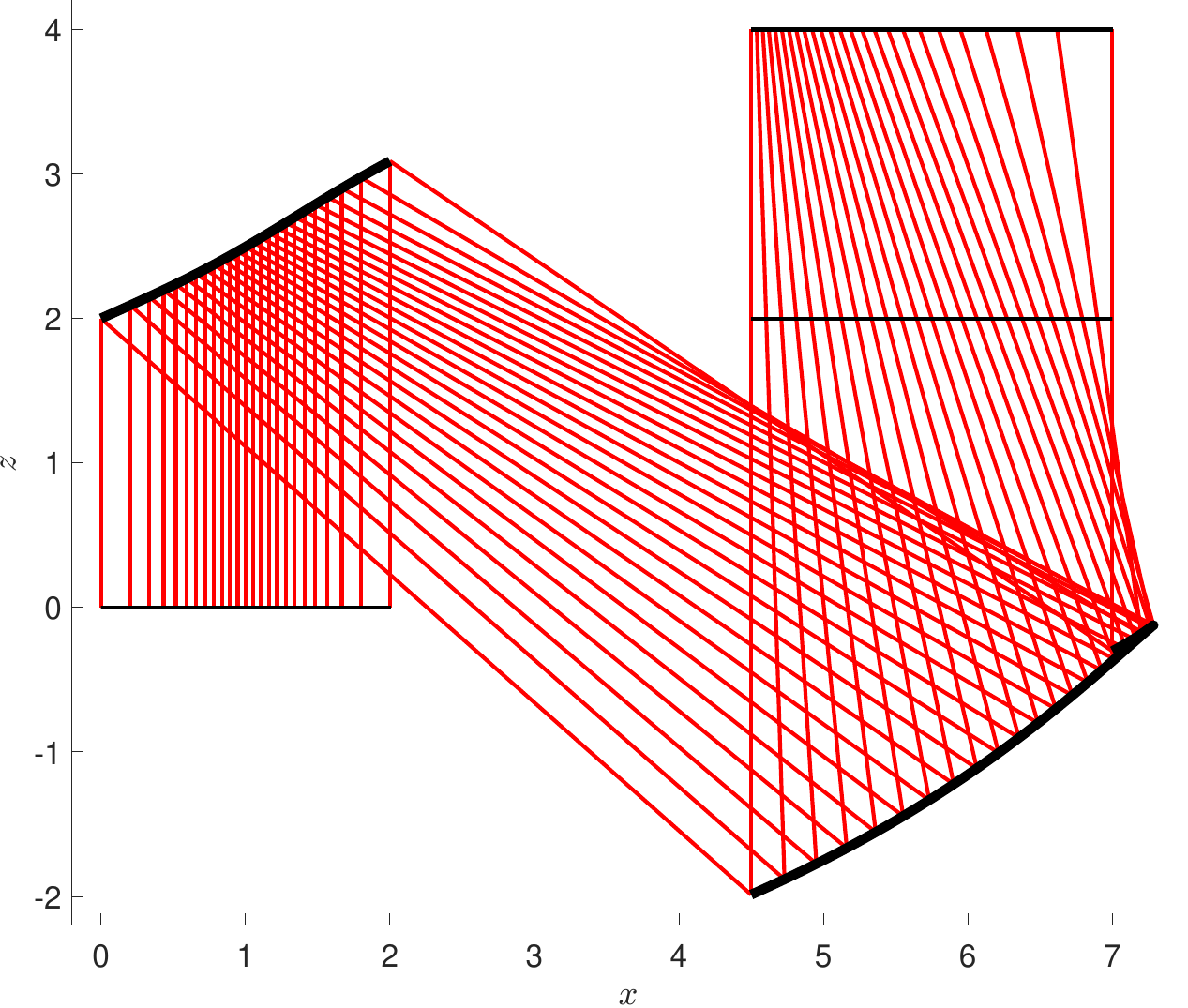}};
    \begin{scope}[x={(image.south east)},y={(image.north west)}]
        \coordinate (spyGlass) at (1.15,0.55);
        \coordinate (spyAt) at (0.91,0.38);
        \spy on (spyAt) in node [] at (spyGlass);
    \end{scope}
    \end{tikzpicture}
    \caption{Self-intersecting reflector.}
    \label{fig:2D_system3}
\end{subfigure}
\caption{Examples of the two-dimensional parallel-to-two-target reflector system.}
\end{figure}

\begin{figure}[!ht]
\centering
\begin{subfigure}[b]{0.48\textwidth}
    \centering
    \includegraphics[width=1.0\textwidth,trim={0cm 0cm 0cm 0cm},clip]{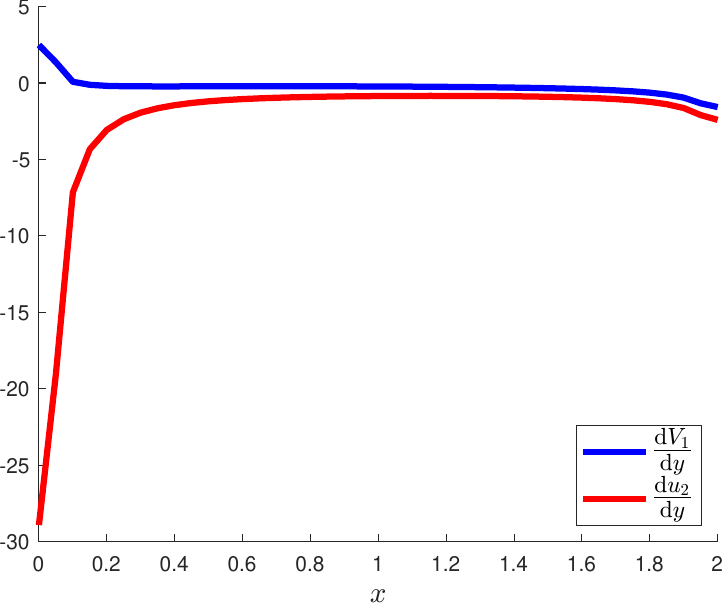}
    \caption{For the reflector system of Fig.~\ref{fig:2D_system1}.}
    \label{fig:1D_R2_0}
\end{subfigure}
\begin{subfigure}[b]{0.48\textwidth}
    \includegraphics[width=1.0\textwidth,trim={0cm 0cm 0cm 0cm},clip]{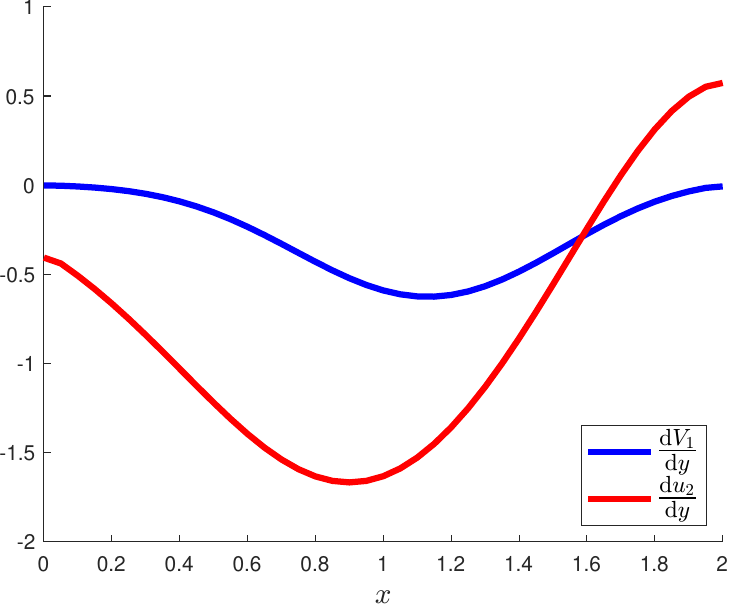}
    \caption{For the reflector system of Fig.~\ref{fig:2D_system3}.}
    \label{fig:1D_R2}
\end{subfigure}
\caption{Derivatives of $V$ and $u_2$ as functions of the source coordinates $x$.}
\end{figure}

As an example, we consider an exponential light distribution $f(x)=\exp{(x-2)}$ on $\mathcal{S}=[0,2]$, a normal distribution $g_1(x)=\exp{((\mu-y)^2/(2\sigma^2))}$ with $\mu=\tfrac{6.5+9}{2}=7.75$ and $\sigma=\sqrt{0.3}$ on $\mathcal{T}_1=[6.5,9]$ and a uniform distribution $g_2(z)=1$ on $\mathcal{T}_2=[7,8]$. We let $L_1=3$, $L_2=4$, set the optical path length $V$ for the left-most light ray to $11.5$ and take $u_1(0)=1.5$. Fig.~\ref{fig:2D_system1} shows the corresponding reflector system with $25$ light rays. The shapes of the reflectors can be verified by ray-tracing. In this example, the reflectors do not self-intersect, which can be verified using the feasibility condition discussed in Sec.~\ref{sec:feasibility}, which states that $\mathcal{R}_2$ does not self-intersect when $u_2'\neq V'$. Fig.~\ref{fig:1D_R2_0} shows a plot of the corresponding functions $u_2'$ and $V'$, which indeed do not intersect and therefore confirm that the reflectors do not self-intersect.

As a second example, we consider a normal light distribution $f(x)=\exp{((\mu-x)^2/(2\sigma^2))}$ with $\mu=1$ and $\sigma=\sqrt{0.3}$ on $\mathcal{S}$, a uniform distribution $g_1(y)=1$ on $\mathcal{T}_1$ and an exponential distribution $g_2(z)=\exp(-z)$ on $\mathcal{T}_2$. Moreover, we let $L_1=2$, $L_2=4$, set the optical path length $V$ for the left-most light ray to $12$ and take $u_1(0)=2$. Fig.~\ref{fig:2D_system3} shows the resulting reflector system with $30$ light rays for these parameters. However, the right end of the second reflector self-intersects, indicating that the prescribed distributions result in a non-physical reflector system. This can be verified using the feasibility condition, for which a plot of the functions $u_2'$ and $V'$ can be seen in Fig.~\ref{fig:1D_R2}. These functions intersect at $x=1.6$ and thereby fail to meet the feasibility condition. To address this, one could, for example, increase the distance between the targets or adjust their domains.

\subsection{Morphing a circle into a parallelogram}
\begin{figure}[!ht]
\centering
\begin{subfigure}[b]{0.3\textwidth}
    \centering
    \includegraphics[width=\textwidth,trim={0cm 0cm 0cm 0cm},clip]{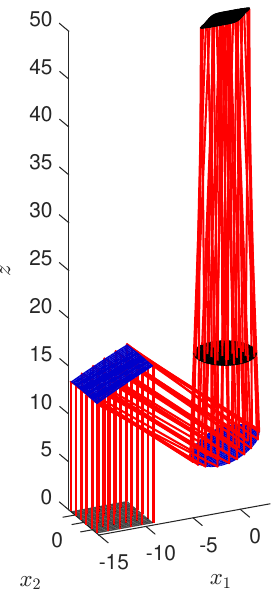}
    \caption{Reflector system.}
    \label{fig:CP_system}
\end{subfigure}
\hfill
\begin{subfigure}[b]{0.65\textwidth}
    \centering
    \begin{subfigure}[b]{0.48\textwidth}
        \centering
        \includegraphics[width=\textwidth,trim={0cm 0cm 0cm 0cm},clip]{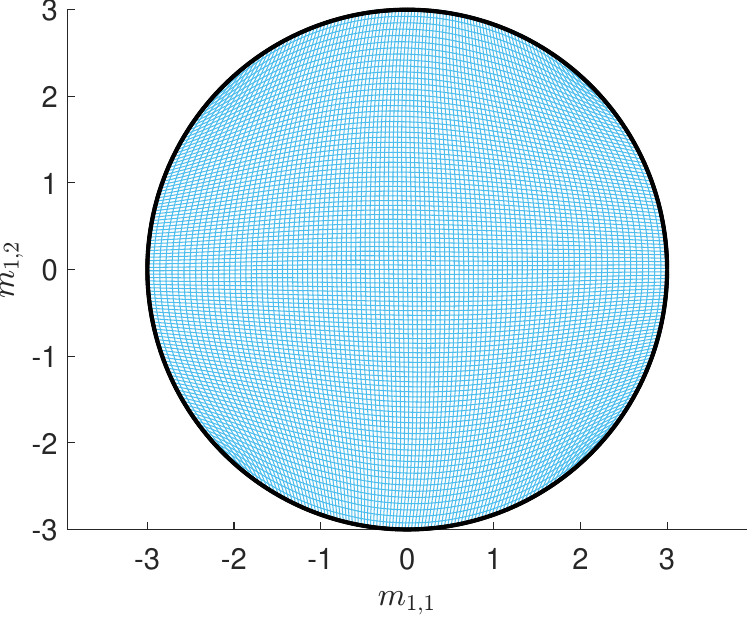}
        \caption{Mapping $\bm{m}_1$ on $\mathcal{T}_1$.}
        \label{fig:CP_m1}
    \end{subfigure}
    \hfill
    \begin{subfigure}[b]{0.48\textwidth}
        \centering
        \includegraphics[width=\textwidth,trim={0cm 0cm 0cm 0cm},clip]{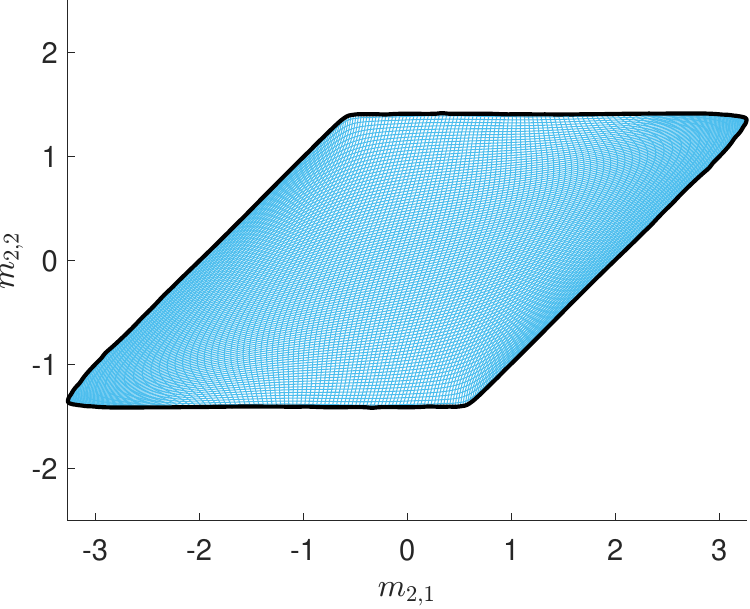}
        \caption{Mapping $\bm{m}_2$ on $\mathcal{T}_2$.}
        \label{fig:CP_m2}
    \end{subfigure}
    \begin{subfigure}[b]{0.48\textwidth}
        \centering
        \includegraphics[width=\textwidth,trim={0cm 0cm 0cm 0cm},clip]{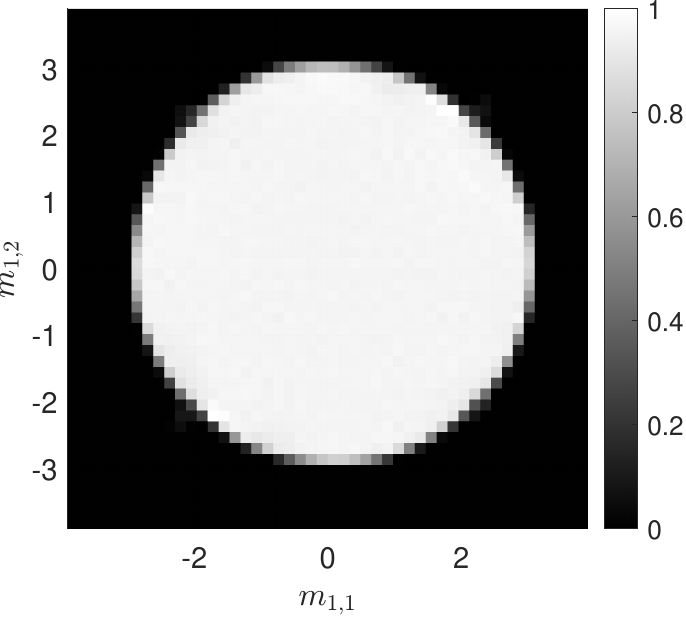}
        \caption{Illuminance pattern on $\mathcal{T}_1$.}
        \label{fig:CP_circle}
    \end{subfigure}
    \hfill
    \begin{subfigure}[b]{0.48\textwidth}
        \centering
        \includegraphics[width=\textwidth,trim={0cm 0cm 0cm 0cm},clip]{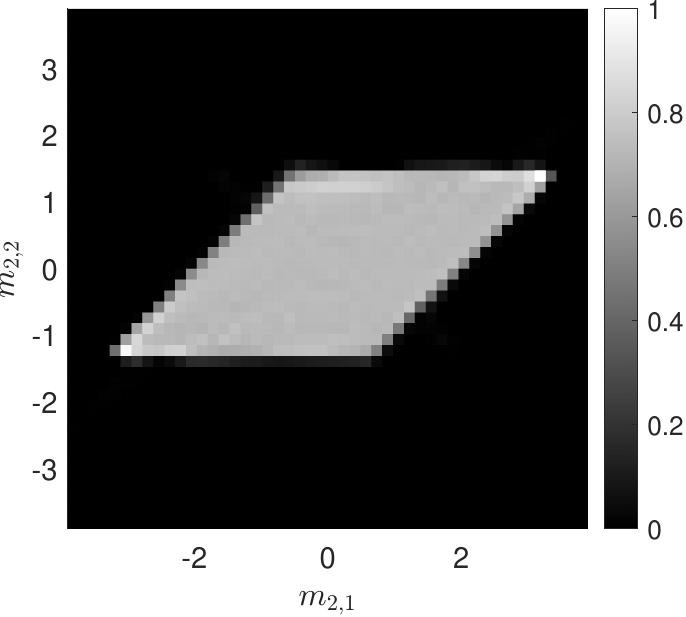}
        \caption{Illuminance pattern on $\mathcal{T}_2$.}
        \label{fig:CP_para}
    \end{subfigure}
\end{subfigure}
\caption{Parallel-to-two-target reflector system with $L_1=15$, $L_2=50$, $V_0=40$, $u_1(\bm{x}_\text{c})=15$, $\alpha=0.5$, $10^4$ iterations of the least-squares algorithm on a $101\times101$ grid.}
\label{fig:ReflectorResults}
\end{figure}

We consider the three-dimensional reflector system with source domain $(x_1,x_2)\in[-15,-9]\times [-3,3]$, first target domain given by the circle centered at $(0,0,L_1)$ with radius $3$ on $z=L_1=15$ and second target domain given by the parallelogram with base length $4$, side length $4$, sharp angle $\tfrac{\pi}{4}$ and intersection of the diagonals $(0,0,L_2)$ on $z=L_2=50$. We consider a uniform light emittance and uniform target illuminances, choose $V(\bm{x}_{\text{c}})=40$ and $u_1(\bm{x}_{\text{c}})=15$ for the center point of the source domain $\bm{x}_{\text{c}}=(-12,0)$, set $\alpha=0.5$ and evaluate $10^4$ iterations of the least-squares algorithm on a $101\times101$ grid, which gives the reflector system in Fig. \ref{fig:CP_system}, where $49$ light rays are shown. In addition, the least-squares algorithm provides the mappings at the two targets, see Figs. \ref{fig:CP_m1} and \ref{fig:CP_m2}, in which we recognize the circle and parallelogram. We verified that the resulting reflector system satisfies the feasibility condition. To verify the reflector shapes, we use a self-programmed quasi-Monte Carlo ray-tracing algorithm in Matlab using $10^6$ rays, which provides us with the three-dimensional irradiance distributions given in Figs. \ref{fig:CP_circle} and \ref{fig:CP_para}, which clearly resemble the uniform distributions of the circle and parallelogram. Of these, $99.56$ percent reaches $\mathcal{T}_1$ in its prescribed domain and $99.98$ percent reaches $\mathcal{T}_2$ in its prescribed domain. Moreover, the root mean squared error of the difference between the expected and actual flux values is $6.55\cdot10^{-7}$ for $\mathcal{T}_1$ and $1.65\cdot10^{-5}$ for $\mathcal{T}_2$.

\subsection{The birth of a chick}\label{sec:chick_egg}
\begin{figure}[!ht]
\centering
\begin{subfigure}[b]{0.305\textwidth}
    \centering
    \includegraphics[width=0.85\textwidth,trim={0cm 0cm 0cm 0cm},clip]{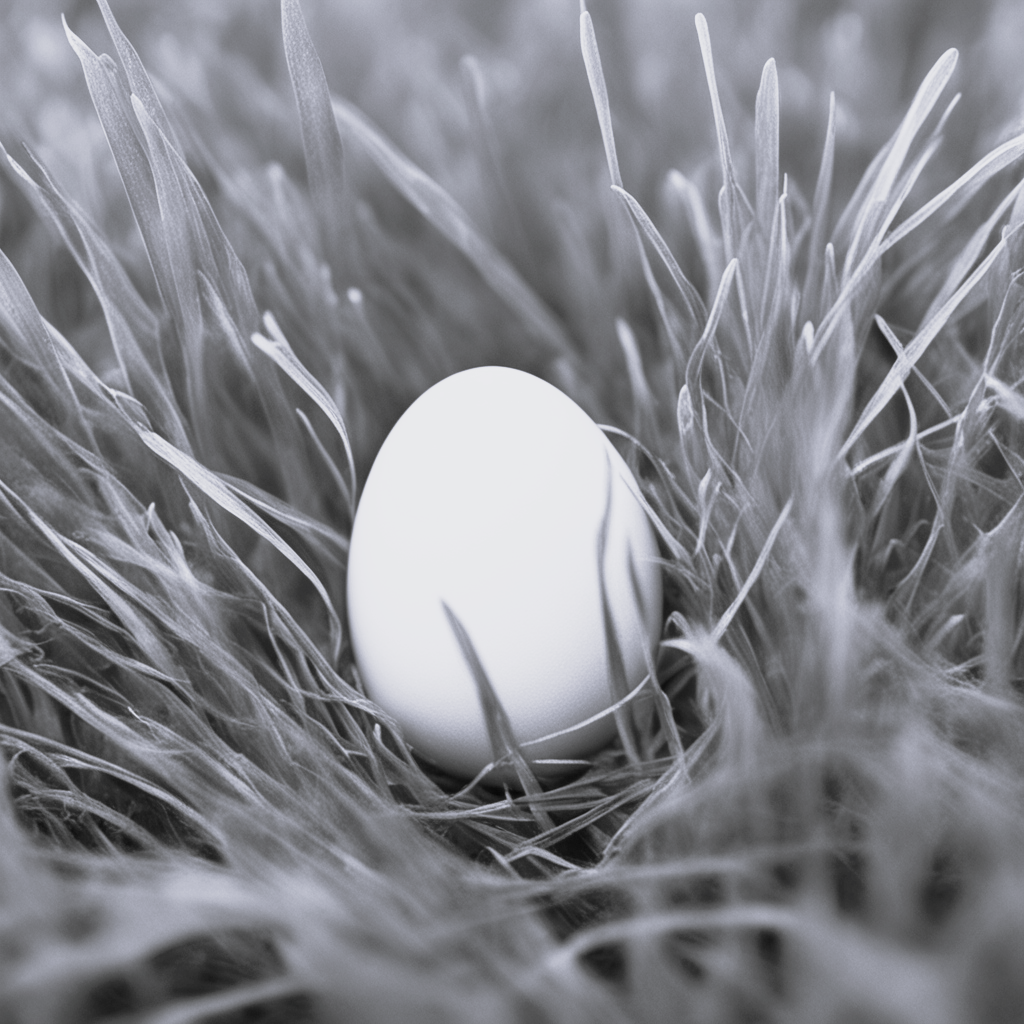}
    \vspace{0.44cm}
    \caption{Distribution pattern of an egg.}
    \label{fig:egg1}
\end{subfigure}
\begin{subfigure}[b]{0.325\textwidth}
    \centering
    \includegraphics[width=0.9\textwidth,trim={0cm 0cm 0cm 0cm},clip]{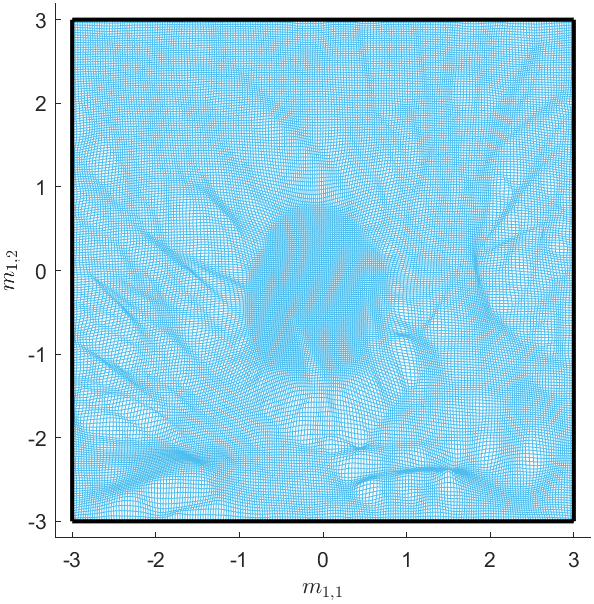}
    \caption{Mapping $\bm{y}=\bm{m}_1(\bm{x})$ on $\mathcal{T}_1$.}
    \label{fig:egg2}
\end{subfigure}
\begin{subfigure}[b]{0.345\textwidth}
    \centering
    \includegraphics[width=0.915\textwidth,trim={0cm 0cm 0cm 0cm},clip]{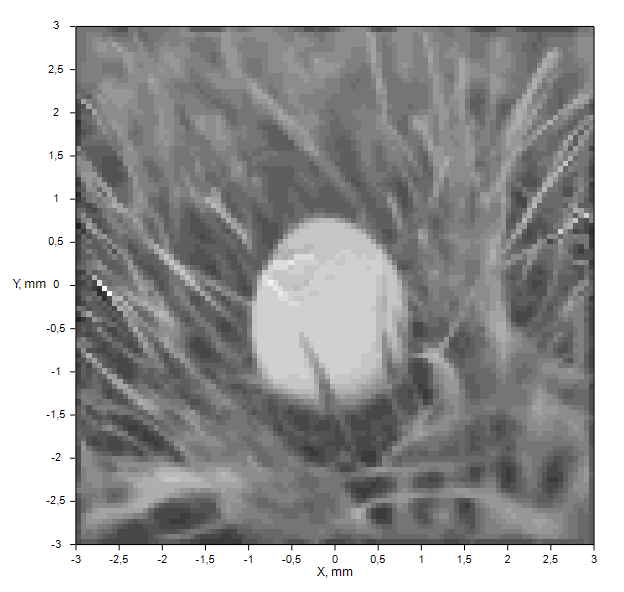}
    \vspace{0.12cm}
    \caption{Ray-traced illuminance pattern on $\mathcal{T}_1$.}
    \label{fig:egg3}
\end{subfigure}
\begin{subfigure}[b]{0.305\textwidth}
    \centering
    \includegraphics[width=0.85\textwidth,trim={0cm 0cm 0cm 0cm},clip]{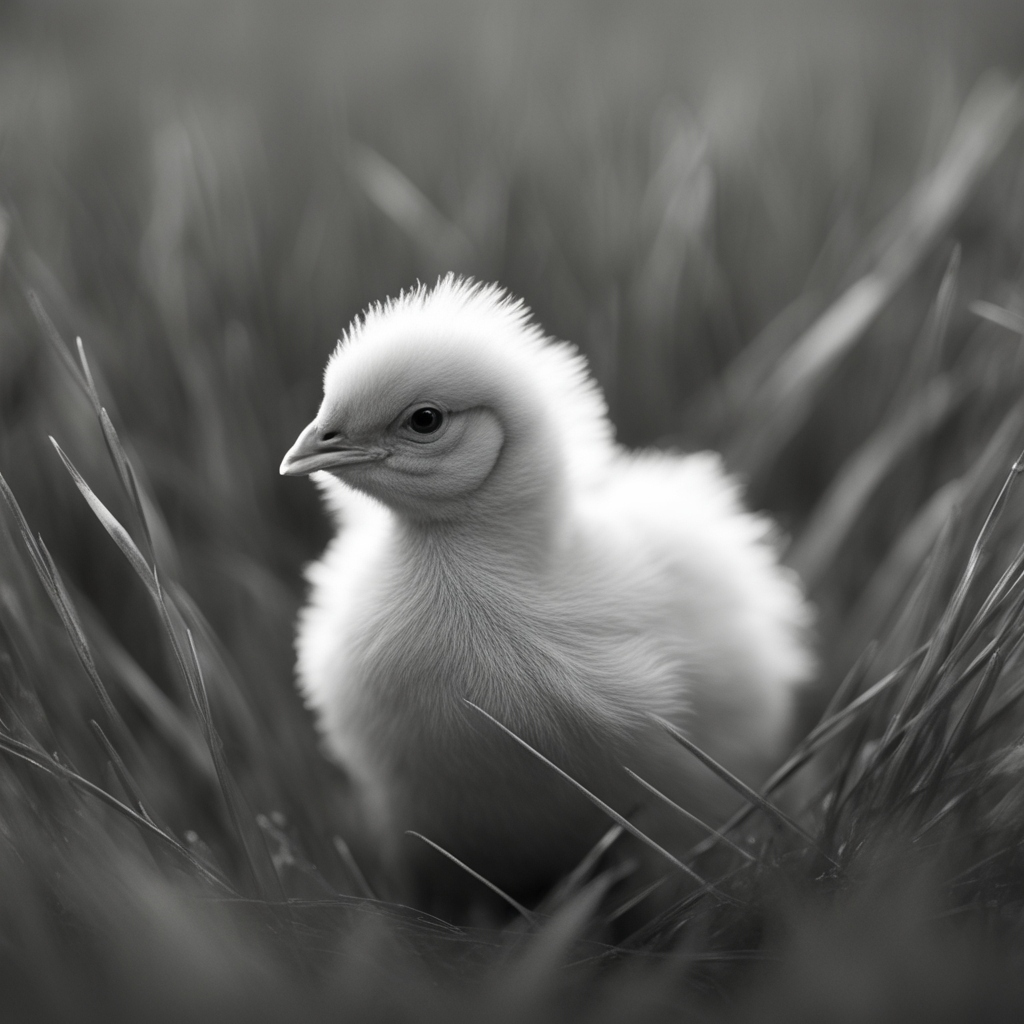}
    \vspace{0.44cm}
    \caption{Distribution pattern of a chick.}
    \label{fig:chick1}
\end{subfigure}
\begin{subfigure}[b]{0.325\textwidth}
    \centering
    \includegraphics[width=0.9\textwidth,trim={0cm 0cm 0cm 0cm},clip]{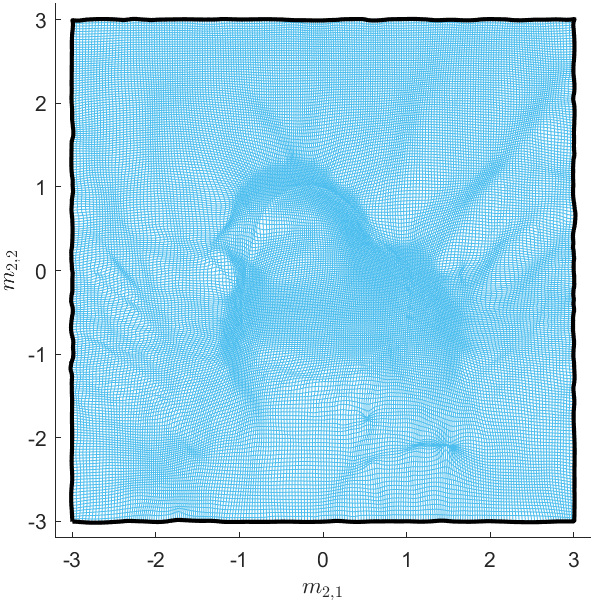}
    \caption{Mapping $\bm{z}=\bm{m}_2(\bm{y})$ on $\mathcal{T}_2$.}
    \label{fig:chick2}
\end{subfigure}
\begin{subfigure}[b]{0.345\textwidth}
    \centering
    \includegraphics[width=0.915\textwidth,trim={0cm 0cm 0cm 0cm},clip]{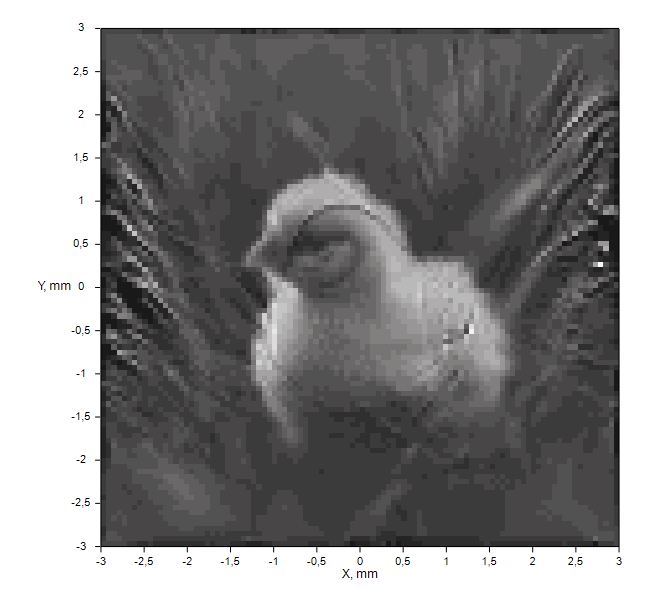}
    \vspace{0.12cm}
    \caption{Ray-traced illuminance pattern on $\mathcal{T}_2$.}
    \label{fig:chick3}
\end{subfigure}
\begin{subfigure}[b]{1.0\textwidth}
    \centering
    \includegraphics[width=0.6\textwidth,trim={0cm 0cm 0cm 0cm},clip]{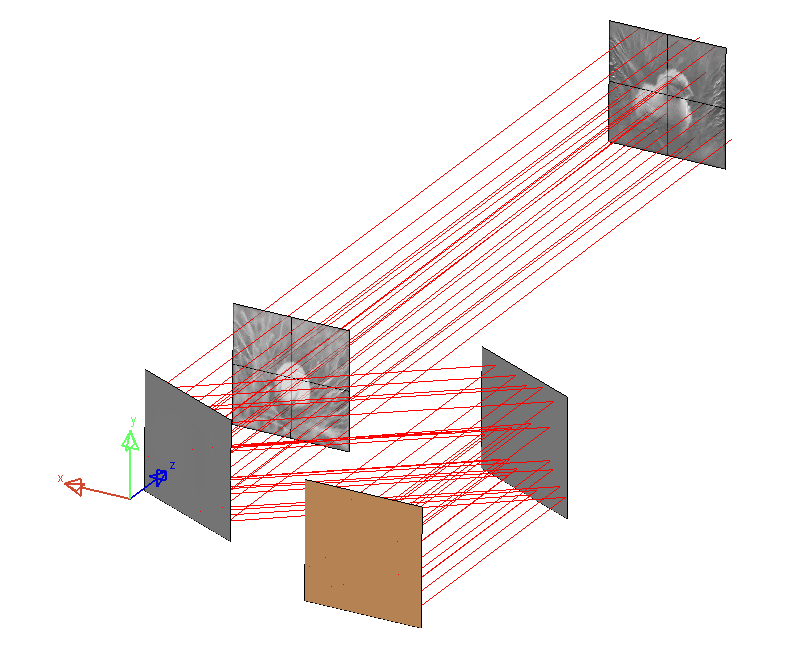}
    \caption{Ray-trace verification in LightTools.}
    \label{fig:3D2}
\end{subfigure}
\caption{Parallel-to-two-target reflector system with $L_1=15$, $L_2=50$, $V_0=40$, $u_1(\bm{x}_\text{c})=15$, $\alpha=0.5$, $10^4$ iterations of the least-squares algorithm on a $201\times201$ grid.}
\label{fig:PawnQueenExample}
\end{figure}

We consider the three-dimensional reflector system with the domains
\begin{align*}
    (x_1,x_2)\in[-15,-9]\times [-3,3],&&
    (y_1,y_2)\in[-3,3]^2,&&
    (z_1,z_2)\in[-3,3]^2.
\end{align*}
The light emittance at the source $\mathcal{S}$ is uniform and the target illuminances at $\mathcal{T}_1$ and $\mathcal{T}_2$ correspond to distribution patterns of an egg and a chick, as shown in Figs. \ref{fig:egg1} and \ref{fig:chick1}. We choose $V(\bm{x}_{\text{c}})=40$ and $u_1(\bm{x}_{\text{c}})=15$ for the center point of the source domain $\bm{x}_{\text{c}}=(-12,0)$. Moreover, we set $L_1=15$, $L_2=50$ and $\alpha=0.5$ and evaluate $10^4$ iterations of the least-squares algorithm on a $201\times201$ grid. The least-squares algorithm provides the mappings at the two targets given in Figs. \ref{fig:egg2} and \ref{fig:chick2}. In these figures, the denser areas indicate brighter regions, from which the distribution patterns of the egg and the chick can clearly be recognized. 

We verified that the resulting reflector system satisfies the feasibility condition and to verify the shapes of the reflectors, we used the commercial ray-tracing software LightTools. We obtained the reflector system in Fig.~\ref{fig:3D2}, which shows 20 light rays originating from the brown-colored source, hitting the two gray-colored reflectors before hitting the two targets. The illuminance patterns on these two targets are illustrated in Figs. \ref{fig:egg3} and \ref{fig:chick3}, which clearly display distribution patterns of the egg and the chick, respectively. Since the least-squares algorithm assumes smooth target distributions, which is not the case in this example, the silhouettes of each distribution pattern can still be faintly seen in the other. This interference can be minimized by using smoother target distributions, by using more grid points or by using an adaptive grid.

\section{Conclusions}\label{sec:conclusion}
This paper presents an inverse method for transforming a given parallel light emittance to two illuminance patterns at different target planes using two freeform reflectors. We modeled the system using the optical path length, generating functions and used energy conservation to find generated Jacobian equations. We derived a feasibility condition that indicates the occurrence of a self-intersecting reflector and verified this condition by considering a two-dimensional simplification of the system. Finally, we used a three-stage least-squares algorithm to compute the three-dimensional reflector system with two complicated target distributions. Overall, the results advance our understanding of how the optical path length can be used to control both the position and momentum of a light ray at the target using an inverse method. In future research we will extend the algorithm by also taking non-parallel sources into account.

\appendix

\begin{backmatter}
\bmsection{Funding}
This work was supported by the NWO Perspectief program \textit{Optical coherence; optimal delivery and positioning} (P21-20).

\bmsection{Disclosures}
The authors declare no conflicts of interest.

\bmsection{Data availability} Data underlying the results presented in this paper are not publicly available at this time but may be obtained from the authors upon reasonable request.
\end{backmatter}



\end{document}